\documentclass[11pt]{article}
\newcommand{\be}{\begin{equation}}
\newcommand{\ee}{\end{equation}}
\newcommand{\ra}{\rightarrow}

\begin{document}

\title{On the Global Anisotropy of Cosmic Ray Data above $4 \times
10^{19}$ eV} \author{{\bf Soebur Razzaque$^1$ and John P. Ralston$^2$} 
\\ \\
$^1$Department of Astronomy and Astrophysics \\ 
Pennsylvania State University, University Park, PA 16802, USA \\ 
$^2$Department of Physics and Astronomy \\ University of Kansas,
Lawrence, KS 66045, USA} \date{} \maketitle

\begin{abstract}
The distribution of arrival directions of ultra-high energy cosmic
rays may yield clues to their mysterious origin.  We introduce a
method of {\it invariant statistics} to analyze cosmic ray data which
eliminates coordinate-dependent artifacts.  When combined with maximum
likelihood analysis, the method is capable of quantifying deviations
of the distribution from isotropy with high reliability.  We test our
method against published AGASA events with energies above $4 \times
10^{19}$ eV.  Angular cuts from observational limitations are taken
into account.  A model based on the Fisher distribution reveals the
rotation of the Earth with the axis $\hat n$ along the direction
($5^h\;53.36^m,\; 85.78^{\circ}$) in $(RA,DEC)$ coordinates, which is
within $5^{\circ}$ of the equatorial north pole. Global anisotropy of
the data, if any, hinges on finer understanding of detector acceptance
than what is available from the published literature.
\end{abstract}

\subsection*{Introduction} 
A puzzle has existed for more than 30 years regarding cosmic ray
events with energies exceeding $4 \times 10^{19}$ eV, a value in the
range of the Greisen, Zatsepin and Kuzmin (GZK) bound \cite{gzk1,
gzk2}.  The nature of the primary particles causing these events is
controversial (see e.g. Refs. \cite{nagano-watson00, biermann-sigl02}
for recent review of the field).  If the primaries are protons from
cosmological distances, then their sources should be isotropically
distributed.  The scrambling effects of intervening magnetic fields
are difficult to assess, but on very basic physics the magnitudes of
the fields (or associated correlation lengths) would have to be
exceedingly small to avoid isotropization.  Diffuse propagation of
cosmic rays from a few prominent sources may result in isotropic
arrival directions as suggested in Refs. \cite{lemoine99, farrar99,
anchordoqui02}.  Cosmic rays from our own Galaxy core are not
considered a viable explanation for the events approaching the GZK
limit \cite{waxman97}.  If our Galaxy substantially modifies
propagation or contributes to production of the primaries above $4
\times 10^{19}$ eV, then any consequent anisotropy should be
correlated with known source(s) of the Galaxy.  On several bases, and
in particular on the basis of {\it symmetry}, one asks whether the
highest energy events are {\it isotropic}, up to the cuts imposed by
observational limitations.

Several studies \cite{uchihori00, hayashida99, takeda99, bird98} at
much lower energies have demonstrated a significant isotropy in
arrival directions.  The AGASA group has presented an analysis of a
sizable (4$\%$) anisotropy of several hundred thousand events with
energies above $10^{17}$ eV.  The study uses right ascension,
declination $(RA, DEC)$ coordinates \cite{hayashida99} and finds a
statistically significant first Fourier moment in $\sin(RA)$.  The
definition of right ascension however, involves historical human
conventions and a choice of axis based on the Earth's orbit.  The
coordinate system axis defining $RA$ requires two parameters for
specification, which are implicitly used in the analysis, potentially
affecting the claimed statistical significance of effects.  In
addition, $\sin \phi$ is not a particular sensitive or informative
statistic, potentially diminishing the impact of 216,000 events.

Here we examine all available data from the AGASA experiment with
energy above $4\times 10^{19}$ eV.  We take care to define certain
quantities of the analysis which previously have been used in a
coordinate dependent way.  Since the coordinate system is a human
artifact, this is to be avoided, unless there is a preferred
coordinate system based on other information.  Our methods are taken
from Refs.  \cite{virmani00} and a large literature on invariant
angular statistics.  Recently Sommers \cite{sommers01} has advocated
an approach using the same basic concepts.

Statistical comparisons require formulation of a well defined {\it
null hypothesis}.  We take particular care on this point.  The term
``null'' often indicates conditions of ``no signal'', which would be
appropriate if one considers anisotropy a signal.  More deeply, the
scientific method proceeds by ruling out possibilities, rather than
attempting to prove particular notions.  The symmetries of data are
the simplest and purest characterizations capable of being tested, or
ruled out.  Thus a solid null hypothesis is the foundation of any
further claims.

The elimination of coordinate dependent artifacts, the testing of a
symmetry-based null, and the objectivity of maximum likelihood
analysis create very powerful tools.  If this method is valid, then it
should be able to determine the systematic bias in real data
objectively.  The main purpose of the paper is to demonstrate the
power of the tools for future use.  However we also test our method to
find significant evidence for anisotropy in the distribution of the
highest energy events, which has been a question of intrinsic interest
for many years.

\subsection*{Methodology} 
The method of maximum likelihood (see Refs.  \cite{kendal1, rpp00} for
example) allows one to test objectively between different statistical
models.  It is good practice to test models which are continuously
related to the null when parameters are varied.  Continuity in the
likelihood test tends to assure that the same features of the data are
tested by the null and competing models.

Let $\vec x$ denote an element of a data set, for example the
coordinates of an observed track.  Given a normalized distribution
$f({\vec x})$, the likelihood of the data in the distribution is
defined to be the product of the distribution evaluated at each data
point over the data set.  Maximum likelihood occurs at maximum log
likelihood ${\cal L}$, defined as
\be
{\cal L}=\sum_{j}^{N} log[f({\vec x_j})] ,
\label{eq:like1}
\ee 
where $N$ is the number of data points.

Hypothesis testing is done by comparing the difference of maximized
log-likelihoods ($T$) for a given data set evaluated with two
different distributions $f({\vec x}), \; f_{\mathrm{null}}({\vec x})$,
where $f_{\mathrm{null}}({\vec x})$ is the null distribution.  For
large $N \gg 1$, the statistic $2T=2({\cal
L}_{\mathrm{model}}^{\mathrm{max}} - {\cal
L}_{\mathrm{null}}^{\mathrm{max}})$ has a chi-squared distribution for
$p$ parameters ($\chi^2_p$).  We will verify this and also determine
the distribution independently by Monte Carlo.  Note that it is not
possible and not our objective to find the ``exact'' distribution
behind the observed data.  Instead, emphasis lies on testing the null
distribution.  It is sufficient to show that a trial distribution fits
the data sufficiently well to rule out the null; one does not imply,
or conclude from this that the trial distribution is the last word.

\subsection*{Covariant and Invariant Statistics} 
For covariant quantities we map the astronomical coordinates $(DEC,
RA)\rightarrow (\pi/2-\theta, \phi) \rightarrow \hat x$, where $\hat
x$ is a unit vector on the dome of the sky, $$\hat x = (\sin\theta
\cos\phi, \, \sin\theta \sin\phi, \,\cos\theta).$$ Naturally $\hat x $
transforms like a vector when the coordinate system is changed.  The
expansion in spherical harmonics are the same thing: vectors are the
$l=1, m$ representations of spherical harmonics.  By contracting
vector indices it is straightforward to make trial distributions and
statistically valid quantities that are scalars under rotations, and
therefore independent of the coordinate system.

The procedure is very natural, and so simple that we should expand on
alternatives which do not have the same features.  It is very common
in the subject of ``circular statistics'' to see quantities such as
$<\!\!\theta\!\!> =\sum_{i}^{N}\theta_{i}/N,\, \Delta \theta = \sqrt{
<\!\!\theta^{2}\!\!>-<\!\!\theta\!\!>^{2} }$.  Such quantities can be
computed but they are so faulty as to be nearly meaningless.  The
fault is immediately seen by calculating $<\!\!\theta\!\!>$ for an
isotropic distribution on a circle.  The distribution has no preferred
orientation, yet the average angle $<\!\!\theta\!\!> \ra \pi$ yields a
preferred point, which is unacceptable.

\subsection*{Analysis} 
We apply our method to the set of 58 published tracks from the AGASA
group with energies not less than $ 4\times 10^{19}$ eV (Table 1,
Refs. \cite{hayashida001}).  The average angular resolution of the
track directions is stated to be 1.8 degrees \cite{uchihori00}
containing one sigma ($68\%$) of the events. The tracks come from the
region of polar angle $ 100^{\circ} < \theta < 10^{\circ}$.  This data
set could be, in principle, anisotropic in any direction.

The world's published data consists of some 200 points.  However not
all are available, while resolution, cuts and data quality varies, and
the dangers of combining data from different groups suggest that
restricting the study to a homogeneous set is the logical first step.
We also want to make it clear that objective and highly significant
conclusions can be extracted from carefully constructed tests without
needing particularly large data sets.  Our methods, of course, extend
readily to much larger data sets anticipated from new arrays including
HIRES and AUGER.

The most basic model \cite{batschelet,fisherni} to analyze unimodal
spherical data \{${\hat x} (\theta, \phi)$\} is the {\it Fisher
distribution}:
\be f_{\mathrm{Fisher}}(\theta, \phi)=\frac{\kappa}{4 \pi\sinh \kappa}
\sqrt {1- ({\hat n}\cdot {\hat x})^2}\,e^{\kappa\,({\hat n}\cdot {\hat
x})} \label{eq:fisher1}
\ee 
where ${\hat n(\alpha, \beta)}$ is the direction of symmetry-breaking
axis.  This distribution has a long history as the spherical
generalization of the venerated von-Mises distribution on the circle.
Both distributions have the analytic properties of the Gaussian
distribution, as seen by writing $$ e^{-\kappa (\hat n-\hat x)^{2}/2 }
\sim e^{\kappa\;(\hat n\cdot \hat x)}, \: \:\hat n^{2}= \hat
x^{2}=1,$$ suppressing normalization factors.  Parameter $\kappa$ is
called the {\it concentration parameter}, and determines the extent
that data is concentrated along the {\it anisotropy axis} $\hat n$.

The Fisher distribution has the minimal number of parameters possible
on the sphere, and exhibits cylindrical symmetry about the axis ${\hat
n}$.  The pre-factor $\sqrt {1- ({\hat n}\cdot {\hat x})^2}$ is
explained as follows.  Choose axis direction $\hat n =$ (0, 0, 1) or
along the {\it north pole }.  This hides 2 parameters, and reduces the
distribution to the simpler form: \be f_{\mathrm{Fisher}}(\theta,
\phi)=\frac{\kappa}{4 \pi\sinh \kappa}\sin\theta e^{\kappa
\cos\theta}. \label{eq:fisher2} \ee

The pre-factor $\sin\theta$ takes into account the Jacobian of solid
angle on the sphere.  Cylindrical symmetry about $\hat n$ is also
obvious now, since $\phi$ does not appear anywhere in
(\ref{eq:fisher2}). For the limit of small anisotropy $\kappa \ll 1$,
the exponential can be expanded $$ e^{\kappa \cos\theta}\sim 1 -
\kappa \cos\theta,$$ which is one of the distributions discussed by
Sommers \cite{sommers01}, and which needs the additional Jacobian
factor to test anisotropy.

The data set is subject to a cut in declination, ($10^{\circ} \le
\theta \le 100^{\circ}$).  We modify the {\it Fisher distribution} to
take this cut into account as:
\be f_{\mathrm{Fisher}}^{\mathrm{cut}}(\theta, \phi) =
\frac{f_{\mathrm{Fisher}}(\theta, \phi) g(\theta)} {\int
{f_{\mathrm{Fisher}}(\theta, \phi) g(\theta) d\theta d\phi}} ;
\label{eq:fishercut} 
\ee 
\be g(\theta) = [\eta (\theta - 10^{\circ})-\eta (\theta -
100^{\circ})],
\label{eq:cut}
\ee
where $\eta$ is the Heaviside step function. The step functions have
an invariant representation we do not bother to write here.  Now
(\ref{eq:fishercut}) is the normalized {\it Fisher distribution} in
the cut region and serves as a trial model of anisotropy.  A variant
of the Fisher distribution known as Watson distribution is useful in
cases where the data is in either bipolar or girdle form and is given
by
\be f_{\mathrm{Watson}}(\theta, \phi)= \frac{\sqrt {1- ({\hat n}\cdot
{\hat x})^2} \,e^{\kappa\;({\hat n}\cdot {\hat x})^2}}{4\pi \int_0^1
e^{-\kappa u^2}du} \label{eq:watson1} \ee
which can be modified to take in to account the cuts in the data in
the same way in (\ref{eq:cut}).

\begin{figure}
\vskip 9.cm \center \begin{picture}(0,0)
\includegraphics{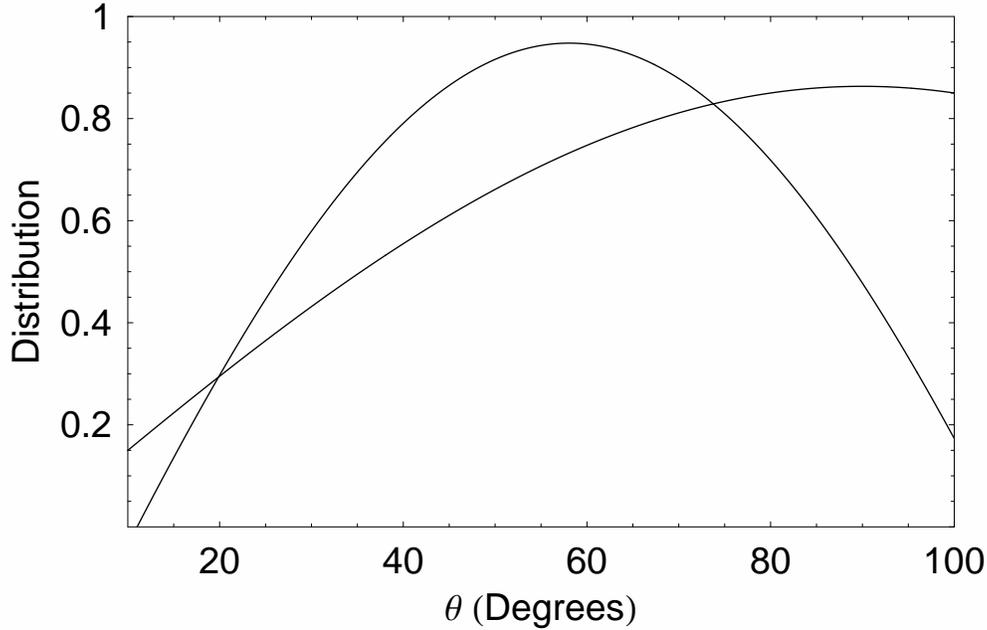}
\end{picture}
\vskip -1.cm \caption{\label{fig:decdist} \small An isotropic
distribution in $\theta$ given in (\ref{eq:null}) and the detector
response observed by AGASA. While the isotropic distribution peaks at
$\theta=90^{\circ}$, the detector response shows a peak at around
$\theta=57^{\circ}$. }
\end{figure}

We also need a {\it null distribution} on the cut region to test our
model using the method of maximum likelihood.  The {\it null isotropic
distribution} on the sphere with the cut in angle $\theta$ is given
by,
\be f_{\mathrm{null}}^{\mathrm{cut}}(\theta, \phi) = \frac{\sin \theta
\;g(\theta)} {\int {\sin \theta \;g(\theta) d\theta d\phi}}.
\label{eq:null}
\ee
The null distribution coincides with
$f_{\mathrm{Fisher}}^{\mathrm{cut}}(\theta, \phi)$ in the limit
$\kappa \ra 0$. Thus when likelihood is evaluated, the variation of
parameter $\kappa$ allows the analysis to be continuously connected
with the null.  However, the detectors usually do not have uniform
acceptance in declination.  AGASA \cite{takeda99} gives a distribution
in declination of all events above $>10^{19}$ eV.  We fit the
distribution by a function of the form ($a \sin b\theta+c$) which is
related to the isotropic distribution in $\theta$, namely $d (\cos
\theta)$.  While the isotropic distribution peaks at
$\theta=90^{\circ}$, the detector response drops as one move away from
$\theta=57^{\circ}$.  The distributions are plotted in Fig.
\ref{fig:decdist}.

\subsection*{Results}
Evaluation of likelihood for the AGASA data set is straightforward.
One finds a maximum log likelihood in the Fisher distribution to be $$
{\cal L}_{\mathrm{Fisher}}=-128.10$$ units at
$$\kappa=1.35 \pm 0.41; \,{\hat n}= (4.22^{\circ}, 88.34^{\circ}).$$ 
In equatorial coordinates ($RA$, $DEC$), the axis of concentration 
lies along the direction ($5^h\;53.36^m,\; 85.78^{\circ}$).  We have 
also evaluated the maximum log likelihood for Watson distribution 
(\ref{eq:watson1}) and the result is: 
${\cal L}_{\mathrm{Watson}}=-129.71$ units.

The null hypothesis is evaluated by running the same analysis with our
fit function to the detector response in $\theta$ shown in
Fig. \ref{fig:decdist}.  The log likelihood is found to be $${\cal
L}_{\mathrm{null}}=-128.81$$ units.  The quantity
$2T^{\mathrm{max}}=2( {\cal L}_{\mathrm{model}}-{\cal
L}_{\mathrm{null}})$ is 1.42 units in case of Fisher distribution
($2T^{\rm max}_{\rm Fisher}$) and 1.81 units in case of Watson
distribution ($2T^{\rm max}_{\rm Watson}$).  However, if the detector
response is not explicitly known or is not well-defined, one uses the
isotropic null distribution (\ref{eq:null}).  The log likelihood is
${\cal L'}_{\mathrm{null}}=-134.39$ units for isotropic null case.
Correspondingly $2T^{'\rm max}_{\rm Fisher}=12.57$ and $2T^{'\rm
max}_{\rm Watson}=9.36$.

For large $N \gg 1$ it is known that $2T$ is distributed like
$\chi^2_p(2T)$ of $p$ parameters.  In practice 30-40 points is
sufficient for large $N$ to apply, but the question also depends on
the dimensionality of the data.  We have $3$ parameters in our case,
so one should expect a $\chi^2_3$ distribution.  This is a prediction
of pure statistical theory, without detail about the problem at hand,
and something that may be questioned.

\begin{figure}
\vskip 9.cm
\center \begin{picture}(0,0)
\includegraphics{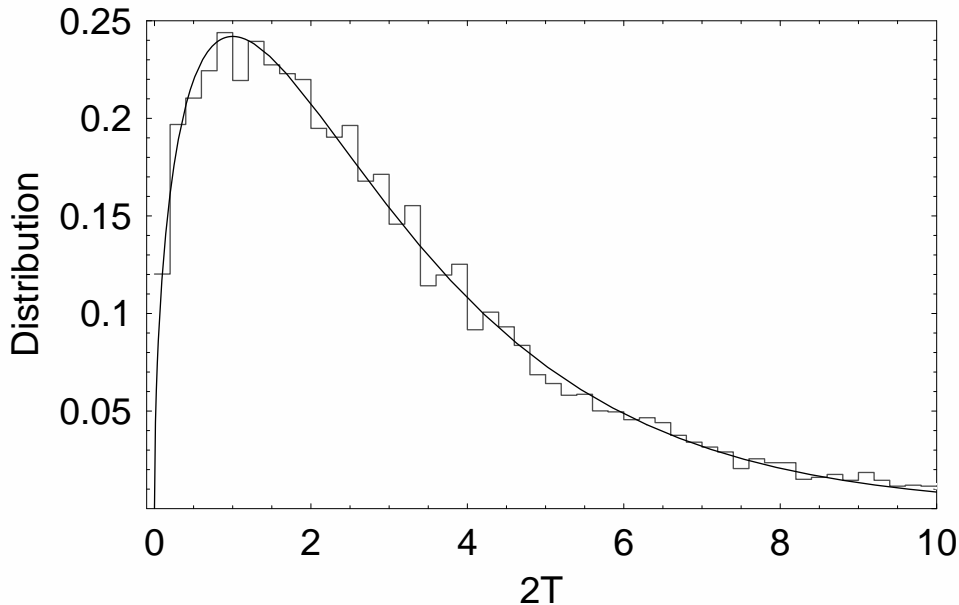}
\end{picture}
\vskip -1.cm
\caption{\label{fig:chi3} 
\small We plot here the $2T$ distribution from Monte Carlo (histogram)
and chi-square distribution of 3 parameters (solid curve). The Monte
Carlo $2T$ distribution was found from ten thousand sets of 58 random
tracks and taking twice the difference between maximum log likelihood
of the Fisher and the isotropic null distributions. }
\end{figure}

Real data sets often contain correlations or irregularities of myriad
possible origin.  Such features may upset analytic estimates.  We
therefore made an independent Monte Carlo study of the statistic $2T$
based on the features of the problem.  We took ten thousand random
samples of $58$ data points inside the cut region.  Randomness was
implemented by selecting points from a distribution flat in $\phi$,
and flat in $\cos\theta$.  For each sample of 58 we varied $\kappa$
and $\hat n$ to determine the maximum likelihood, and we calculated
the likelihood for $\kappa=0$; the result was a trial value of $2T$.
The distribution of $2T$ values for 10,000 runs is shown in Fig
\ref{fig:chi3} along with the $\chi^2_3$ distribution. Agreement is
excellent.

Statistical significance is determined by calculating $P$-values, also
known as confidence levels, which are defined as the integrated
probability of $2T$ to fluctuate in the null distribution above the
value determined for the data.  A comparison of $P$-values is also a
method to determine whether the distributions have any unexpected long
tails.  The $P$-value for the $\chi^2_3$ distribution to give $2T \geq
1.42$ or $2T \geq 1.81$ is very large.  It means the AGASA data set is
rather isotropic and do not strongly point towards a unique source
direction or do not suggest that the data points are concentrated in a
girdle form.  This is the conclusion from using detector response in
declination reported by AGASA group.  On the other hand, from
isotropic null distribution, the $P$-value for the $\chi^2_3$
distribution to give $2T \geq 12.57$ is $3.9 \times 10^{-4}$, or
0.039\%.  In terms of 2-sided Gaussian statistics, which are often
used for comparisons, $P=3.9 \times 10^{-4}$ corresponds to a ``3.55
$\sigma$'' effect.

\subsection*{Observations} 
\begin{figure}
\vskip 9.cm \center \begin{picture}(0,0)
\includegraphics{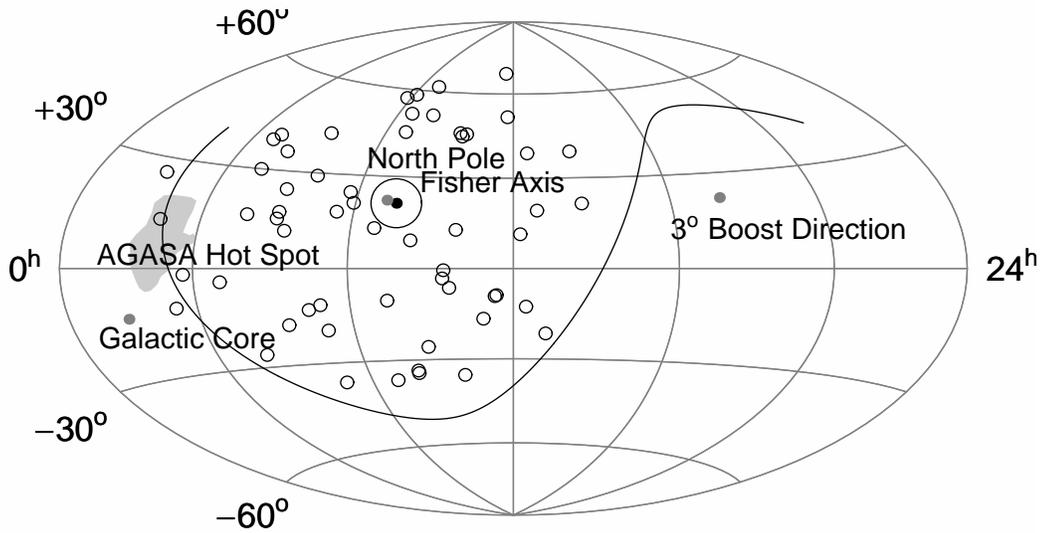}
\end{picture}
\vskip -1.5cm \caption{\label{fig:hammer} 
\small Aitoff-Hammer equal area projection of the sky in Galactic
Coordinates. The AGASA tracks are denoted by open circles with a
radius equal to their angular resolution of 1.8 degrees. The
anisotropic hot spot AGASA group found at lower energies
\cite{hayashida001} has been shown as the shaded region at the left
corner. The Fisher axis of global anisotropic direction we have found
is about $90^{\circ}$ from the AGASA hot spot and close to the
equatorial north pole. The error in the determination of the Fisher
axis is shown by the circle We also show the direction of motion
relative to $3^{\circ}$ cosmic microwave background and the equatorial
plane. }
\end{figure}

We plotted the data in several coordinate systems to visually examine
signals of anisotropy, and to check the fit to the Fisher
distribution.  Results in Galactic Coordinates and Aitoff-Hammer
equal-area projection are shown in Fig \ref{fig:hammer}. The advantage
of an equal-area projection is that an isotropic distribution will
appear uniform, and anisotropy is not unduly distorted by the
coordinates. Galactic coordinates are used so that any correlation
with the galaxy plane might be easily seen. The equatorial plane is
shown by the solid curve.

The figure shows the Fisher ``hot-spot'' extracted from the fit,
namely the best-fit anisotropy axis $\hat n$.  As can be seen, the
Fisher axis is close to the equatorial north pole and its error cone
actually encloses the north pole.  Also shown are the locations of
some other features of obvious interest: the hot-spot of the lower
energy AGASA analysis \cite {takeda99}, and the approximate location
of the galaxy core. We added the direction of our system's motion
relative to the cosmological background radiation, as deduced from the
dipole component of the the $3^{\circ}$ cosmic microwave background,
for another comparison.

\subsection*{Conclusion} 
Unresolved questions regarding the origin of cosmic rays with energies
above $4 \times 10^{19}$ eV suggests testing the degree of isotropy of
the data.  The method we have developed is capable of ruling out
isotropy and eliminating any coordinate dependent artifacts. Given the
detector bias reported by AGASA, the group's cosmic ray tracks above
$4 \times 10^{19}$ eV are consistent with an isotropic
distribution. The question of anisotropy probably hinges on a finer
understanding of angular bias than currently available in published
sources: for example, error bars on the angular acceptance would be
useful.

\vspace{11pt} \noindent {\bf Acknowledgements:} This study springs out
of collaboration with Pankaj Jain, who we thank for productive
comments and criticsm.  Work was supported in part under the
Department of Energy grant number DE-85ER40212 by the University of
Kansas General Research Fund, and the {\it Kansas Institute for
Theoretical and Computational Science/K*STAR} program.

\newcommand{\noopsort}[1]{} \newcommand{\printfirst}[2]{#1}
  \newcommand{\singleletter}[1]{#1} \newcommand{\switchargs}[2]{#2#1}

\begin{center}
\begin{tabular}{lccc|cccc}
\multicolumn{8}{c}{Table 1: AGASA Events Above $4 \times 10^{19}$ eV
(J2000 coordinates)} \\
\hline \hline
Date & Energy & $RA$ & $DEC$ & Date & Energy & $RA$ & $DEC$ \\
     & ($\times 10^{19}$ eV) & & & & ($\times 10^{19}$ eV) & & \\
\hline
84/12/12 & 6.81 & 22$^{h}$ 21$^{m}$ & 38.4$^{\circ}$ &
84/12/17 & 9.79 & 18$^{h}$ 29$^{m}$ & 35.3$^{\circ}$ \\
86/01/05 & 5.47 &  4$^{h}$ 38$^{m}$ & 30.1$^{\circ}$ &
86/10/23 & 6.22 & 14$^{h}$ 02$^{m}$ & 49.9$^{\circ}$ \\
87/11/26 & 4.82 & 21$^{h}$ 57$^{m}$ & 27.6$^{\circ}$ &
89/03/14 & 5.27 & 13$^{h}$ 48$^{m}$ & 34.7$^{\circ}$ \\
89/08/16 & 4.07 &  5$^{h}$ 51$^{m}$ & 58.5$^{\circ}$ &
90/11/25 & 4.51 & 16$^{h}$ 17$^{m}$ & $-$7.2$^{\circ}$ \\
91/04/03 & 5.09 & 15$^{h}$ 47$^{m}$ & 41.0$^{\circ}$ &
91/04/20 & 4.35 & 18$^{h}$ 59$^{m}$ & 47.8$^{\circ}$ \\
91/05/31 & 5.53 &  3$^{h}$ 37$^{m}$ & 69.5$^{\circ}$ &
91/11/29 & 9.10 & 19$^{h}$ 06$^{m}$ & 77.2$^{\circ}$ \\
91/12/10 & 4.24 &  0$^{h}$ 12$^{m}$ & 78.6$^{\circ}$ &
92/01/07 & 4.51 &  9$^{h}$ 36$^{m}$ & 38.6$^{\circ}$ \\
92/01/24 & 4.88 & 17$^{h}$ 52$^{m}$ & 47.9$^{\circ}$ &
92/02/01 & 5.53 &  0$^{h}$ 34$^{m}$ & 17.7$^{\circ}$ \\
92/03/30 & 4.47 & 17$^{h}$ 03$^{m}$ & 31.4$^{\circ}$ &
92/08/01 & 5.50 & 11$^{h}$ 29$^{m}$ & 57.1$^{\circ}$ \\
92/09/13 & 9.25 &  6$^{h}$ 44$^{m}$ & 34.9$^{\circ}$ &
93/01/12 & 10.1 &  8$^{h}$ 17$^{m}$ & 16.8$^{\circ}$ \\
93/01/21 & 4.46 & 13$^{h}$ 55$^{m}$ & 59.8$^{\circ}$ &
93/04/22 & 4.42 &  1$^{h}$ 56$^{m}$ & 29.0$^{\circ}$ \\
93/06/12 & 6.49 &  1$^{h}$ 16$^{m}$ & 50.0$^{\circ}$ &
93/12/03 & 21.3 &  1$^{h}$ 15$^{m}$ & 21.1$^{\circ}$ \\
94/07/06 & 13.4 & 18$^{h}$ 45$^{m}$ & 48.3$^{\circ}$ &
94/07/28 & 4.08 &  4$^{h}$ 56$^{m}$ & 18.0$^{\circ}$ \\
95/01/26 & 7.76 & 11$^{h}$ 14$^{m}$ & 57.6$^{\circ}$ &
95/03/29 & 4.27 & 17$^{h}$ 37$^{m}$ & $-$1.6$^{\circ}$ \\
95/04/04 & 5.79 & 12$^{h}$ 52$^{m}$ & 30.6$^{\circ}$ &
95/10/29 & 5.07 &  1$^{h}$ 14$^{m}$ & 20.0$^{\circ}$ \\
95/11/15 & 4.89 &  4$^{h}$ 41$^{m}$ & 29.9$^{\circ}$ &
96/01/11 & 14.4 & 16$^{h}$ 06$^{m}$ & 23.0$^{\circ}$ \\
96/01/19 & 4.80 &  3$^{h}$ 52$^{m}$ & 27.1$^{\circ}$ &
96/05/13 & 4.78 & 17$^{h}$ 56$^{m}$ & 74.1$^{\circ}$ \\
96/10/06 & 5.68 & 13$^{h}$ 18$^{m}$ & 52.9$^{\circ}$ &
96/10/22 & 10.5 & 19$^{h}$ 54$^{m}$ & 18.7$^{\circ}$ \\
96/11/12 & 7.46 & 21$^{h}$ 37$^{m}$ & 8.1$^{\circ}$  &
96/12/08 & 4.30 & 16$^{h}$ 31$^{m}$ & 34.6$^{\circ}$ \\
96/12/24 & 4.97 & 14$^{h}$ 17$^{m}$ & 37.7$^{\circ}$ &
97/03/03 & 4.39 & 19$^{h}$ 37$^{m}$ & 71.1$^{\circ}$ \\
97/03/30 & 15.0 & 19$^{h}$ 38$^{m}$ & $-$5.8$^{\circ}$ &
97/04/10 & 3.89 & 15$^{h}$ 58$^{m}$ & 23.7$^{\circ}$ \\
97/04/28 & 4.20 &  2$^{h}$ 18$^{m}$ & 13.8$^{\circ}$ &
97/11/20 & 7.21 & 11$^{h}$ 09$^{m}$ & 41.8$^{\circ}$ \\
98/02/06 & 4.11 &  9$^{h}$ 47$^{m}$ & 23.7$^{\circ}$ &
98/03/30 & 6.93 & 17$^{h}$ 16$^{m}$ & 56.3$^{\circ}$ \\
98/04/04 & 5.35 & 11$^{h}$ 13$^{m}$ & 56.0$^{\circ}$ &
98/06/12 & 12.0 & 23$^{h}$16$^{m}$ & 12.3$^{\circ}$ \\
98/09/03 & 4.69 & 19$^{h}$36$^{m}$ & 50.7$^{\circ}$ &
98/10/27 & 6.11 &  3$^{h}$45$^{m}$ & 44.9$^{\circ}$ \\
99/01/22 & 7.53 & 19$^{h}$11$^{m}$ &  5.3$^{\circ}$ &
99/07/22 & 4.09 &  7$^{h}$39$^{m}$ & 32.2$^{\circ}$ \\
99/07/28 & 7.16 &  3$^{h}$46$^{m}$ & 49.5$^{\circ}$ &
99/09/22 & 10.4 & 23$^{h}$03$^{m}$ & 33.9$^{\circ}$ \\
99/09/25 & 4.95 & 22$^{h}$40$^{m}$ & 42.6$^{\circ}$ &
99/10/20 & 6.19 &  4$^{h}$37$^{m}$ &  5.1$^{\circ}$ \\
99/10/20 & 4.29 &  4$^{h}$02$^{m}$ & 51.7$^{\circ}$ &
00/05/26 & 4.98 & 14$^{h}$08$^{m}$ & 37.1$^{\circ}$ \\
\hline \hline
\end{tabular}
\end{center}


\begin{thebibliography}{10}

\bibitem{gzk1} K.~Greisen.  \newblock {\em Phys. Rev. Lett.}, 16:748,
1966.

\bibitem{gzk2} G.~T. Zatsepin and V.~A.~Kuzmin.  \newblock {\em
Sov. Phys. JETP Lett.}, 4:78, 1966.

\bibitem{nagano-watson00} M.~Nagano and A.~A.~Watson. \newblock {\em
Rev. Mod. Phys.}, 72:689, 2000.

\bibitem{biermann-sigl02} P.~Biermann and G.~Sigl. \newblock {\em
Lect. Notes Phys.}, 576:1, 2001 [arXiv:astro-ph/0202425].

\bibitem{lemoine99} M.~Lemoine, G.~Sigl and P.~Biermann.
arXiv:astro-ph/9903124

\bibitem{farrar99} G.~R.~Farrar and T.~Piran. \newblock {\em
Phys. Rev. Lett.}, 84:3527, 2000 [arXiv:astro-ph/9906431].

\bibitem{anchordoqui02} L.~A.~Anchordoqui, H.~Goldberg and
D.~F.~Torres. \newblock arXiv:astro-ph/0209546.

\bibitem{waxman97} E.~Waxman, K.~B.~Fisher and T.~Piran.  \newblock
{\em Astrophys. J.}, 483:1, 1997 [arXiv:astro-ph/9604005].

\bibitem{uchihori00} Y.~Uchihori et~al.  \newblock {\em
Astropart. Phys.}, 13:151, 2000.

\bibitem{hayashida99} N.~Hayashida et~al.  \newblock astro-ph/9906056.

\bibitem{takeda99} M.~Takeda et~al.  \newblock astro-ph/9902239.

\bibitem{bird98} D.~J.~Bird {\it et al.}  [HIRES
Collaboration]. \newblock arXiv:astro-ph/9806096.

\bibitem{virmani00} A.~Virmani, S.~Bhattacharya, P.~Jain, S.~Razzaque,
J.~P. Ralston, and D.~W.  McKay.  \newblock Astropart.\ Phys.\ {\bf
17}, 489 (2002).

\bibitem{sommers01} Paul Sommers.  \newblock {\em Astropart. Phys.},
14:271, 2001.

\bibitem{kendal1} Alan Stuart and J.~Keith Ord.  \newblock {\em
Kendall's Advance Theory of Statistics}, volume~1.  \newblock Halsted
Press, 1994.

\bibitem{rpp00} Particle~Data Group.  \newblock Review of particle
physics.  \newblock {\em The European Physical Journal C}, 15, 2000.

\bibitem{hayashida001} N.~Hayashida et~al.  \newblock {\em
Astrophys. J.}, 522:225, 1999.

\bibitem{batschelet} Edward Batschelet.  \newblock {\em Circular
Statistics in Biology}.  \newblock Academic Press, 1981.

\bibitem{fisherni} N.~I. Fisher, T.~Lewis, and B.~J.~J. Embleton.
\newblock {\em Statistical analysis of spherical data}.  \newblock
Cambridge University Press, 1987.

\end{thebibliography}
\end{document}